\documentclass[aps,prl,twocolumn,superscriptaddress,10pt]{revtex4-2}
\usepackage[dvips]{graphicx}  
\usepackage{natbib}
\usepackage{datetime}
\usepackage{xcolor}
\usepackage{ulem}
\usepackage{epstopdf}
\usepackage{amssymb}
\usepackage{amsmath}
\colorlet{red}{black}
\colorlet{blue}{black}
\def\d{{\rm d}}

\newcommand{\olsi}[1]{\,\overline{\!{#1}}} 
\def\be{\begin{eqnarray}}
\def\ee{\end{eqnarray}}
\def\der{{\rm d}}
\def\e{{\rm e}}
\def\({\left(}
\def\){\right)}

\begin{document}
\title{Size distributions in irreversible particle aggregation}
\author{Klavs Hansen}
\email{KlavsHansen@tju.edu.cn}
\affiliation{Center for Joint Quantum Studies and Department 
of Physics, School of Science, Tianjin University,\\
92 Weijin Road, Tianjin 300072, China}
\date{\today,~\currenttime}

\begin{abstract}
The aggregation of particles in the free molecular regime 
is determined approximately for {\color{blue} situations
with a high degree of translational energy equilibration}.
The mean particle sizes develop linearly in time.
Scaling relations are used to derive a linear partial 
differential equation which is solved to show that the size 
distributions are close to log-normal asymptotically
in time.
\end{abstract}
\maketitle

\section{Motivation}

The growth of particles by irreversible molecular accumulation 
and cluster-cluster aggregation in a closed system in 
quasi-equilibrium is governed by the aggregation Schmoluchowski 
{\color{blue} equations} \cite{SmoluchowskiZPC1917}.
The equations have gained renewed interest with the appearance 
of nanoparticles as technologically interesting species.
With their strongly size-dependent properties 
\cite{Haberland1994}, control and knowledge of size 
distributions and the factors that determine them therefore 
become of prime interest.
A good understanding of the growth of particles is likewise 
highly relevant for a quantitative description of the 
kinetics of atmospheric nucleation, in parallel to  
thermodynamic quasi-equilibrium description{\color{blue}s} 
\cite{HoltenJCP2009,FordJMES2004,ElmJAS2020}.
Similarly, it is essential for {\color{blue} the technological 
applications} where, for example, time-dependent kernels 
have been suggested to engineer 3d printing for medicine 
\cite{MamontovAMM2017}. 

The present work aims {\color{blue} to provide} 
solutions to the equations under the conditions characterized 
as the free molecular regime \cite{AldousBernoulli1999}.
{\color{red} The general approach here} is that 
of mean field theory where, at a given time, a single 
concentration for each particle size describes both the 
state of the system and the growth of the particles. 
This disregards the fluctuations that must be inherent 
in the stochastic processes described by the Smoluchowski 
equations \cite{MarcusTm1968,LushnikovIAN1979}.
On general grounds we must expect fluctuations to influence 
the solutions at most to second order in their relative 
values, although this is not guaranteed (see ref. 
\cite{AldousBernoulli1999} and references therein).
Including fluctuations will have the strongest consequences 
for the low-intensity clusters. 

As has been discussed extensively by Brilliantov and 
collaborators \cite{BrilliantovPRE2020,OsinskiPRE2022}, 
ballistic aggregation implies a time development of the rate 
coefficients (kernels) that enter the equations. 
The choice made here corresponds to a sufficiently large 
scattering cross section of the aggregating particles
{\color{blue} or, equivalently, to a small fraction of 
merging collisions}.
Scattering will occur due to the long range inter-particle
forces, for simplicity assumed to be elastic, following 
\cite{BrilliantovPRE2020}.

{\color{red}The analysis will focus on the bulk part of 
the size distributions, with no attempt to describe 
the extremes {\color{blue} of the distribution,} far 
beyond the average sizes, i.e. the concentrations for 
$N\gg \olsi{N}$ or $N \ll \olsi{N}$.
Although these are of obvious interest for, e.g. size 
distributions of particles in connection with planetary 
growth \cite{Wilkinson2008,BrilliantovPNAS2015}, the 
focus here is on systems of limited volume and 
particle number{\color{blue}s}.
For these, descriptions in terms of homogeneous densities 
are less than rigorous, and {\color{blue} they are} 
some of the systems for which the extreme, low abundance 
parts of the size distributions become most uncertain.
Such systems include as prime examples the production 
of clusters and nanoparticles in sources where aggregation 
takes place in regions limited in both time and space.}

The results from a number of experimental 
studies of cluster production have shown results similar to 
those reported here, dating from the early days of cluster 
science to the present, even if authors do not always 
interpret the experimental results in terms of the 
log-normal distribution derived here{\color{blue}, or
discuss the distributions at all. 
References \cite{HallZPD1989,McHughZPD1989,BjornholmPRL1990,
PedersenNature1991,YeretzianZPD1993,GimelsheinJCP2015,
FischerPRR2022} provide a few examples of apparently 
log-normal distributions for some different clustering 
materials.}

A few different kernels (the $a$'s in Eq.(\ref{smo-gain}) 
below) are known to lend themselves to exact, closed form 
solutions.
Of special interest here is the solution for the 
size-independent kernel used by Smoluchowski, which yields 
a single exponential decay.
It will also be a potential solution to the equations 
here, together with the log-normal distribution.
Which of the two possibilities will be realized will be 
determined in a delicate manner by the form of the kernels.
It should be mentioned that other theoretical works have 
found log-normal distributions, see e.g. ref. 
\cite{BergmannJCG2008}, although the processes described 
are significantly different from {\color{blue} those 
described} in the present work.

\section{Fundamental equation and kernels}

The time development of an irreversibly aggregating particle 
distribution is described by the coupled ordinary differential 
equations
\begin{equation}
\label{smo-gain}
\frac{\der c_N}{\der t}
= \sum_{i =1}^{N-1} \frac{a_{i,N-i}}{2} c_i c_{N-i} 
- \sum_{i=1}^{\infty} a_{i,N} c_N c_i,
\end{equation}
where $c_i$ denotes the concentrations of particle size $i$. 
{\color{red}The first term describes the formation of particle 
size $N$ by fusion of two particles of sizes $N-i$ and $i$. 
Ranging over all values of $i$ requires the factor 1/2 to 
compensate for double-counting.
The second term describes the losses due to the formation of 
larger particles from size $N$.}
The kernels, $a_{i,j}$, are defined by the physical situation, 
but we can assume without any further justification that they 
are symmetric in their indices, $a_{i,j} = a_{j,i}$.

{\color{red}The situation described by the equations is 
somewhat idealized. 
It will for example not consider the bottleneck caused by
the need for a three-body collision for creating a stable 
dimer in the absence of internal degrees of freedom that 
can be excited, i.e., for nucleation of atomic species
at ambient conditions.
The equations do not include bouncing collisions, 
fragmentation{\color{blue}, or} shattering.
The collision energies are assumed to be so low that these 
processes do not happen.
Low energy collisions could potentially be non-sticking in 
the presence of an attachment barrier.
This possibility will be disregarded here, as such barriers 
are rarely seen outside the realm of chemical bond 
formation.
Finally, we note that the requirement of elastic scattering
collisions must be relaxed during the initial 
phases of aggregation, in order to carry away the heat of 
formation of the particles. 
At later stages, the surface energies provide the heat of 
fusion, and this varies as the monomer binding energy 
with size, with $N^{2/3}$ to be specific.
The evaporation rates are essentially determined by 
the monomer binding energy divided by the energy per degree 
of freedom and the parallel variations of the heating and 
stability will therefore not cause an upper limit to the 
application of the equations here.}

The equations obey total particle number conservation.
It follows from a rewrite of the first term {\color{red} on 
the right-hand side of the equation.
The summations over $i$ are diagonal in an $N,i$ diagram and
the summation over $N$ adds up these to cover the whole plane.
The double sum can be resolved into a summation along the 
$N$-axis and one along the $i$-axis, with coefficients $N$ 
and $i$:}
\be
&&\sum_{N=1}^{\infty} N \sum_{i =1}^{N-1} \frac{a_{i,N-i}}{2} 
c_i c_{N-i}\\\nonumber
&=&\frac{1}{2}\(\sum_{N=1}^{\infty}{\color{red}N}
\sum_{i=1}^{\infty}a_{i,N}c_ic_N  
+\sum_{i=1}^{\infty}i\sum_{N=1}^{\infty} a_{N,i} 
c_{\color{red}N} c_{\color{red}i}\),
\ee
which cancels the last sum in the expression.

The choice of the physical situation gives the following 
kernels:
\be
a_{i,j} = \sigma_{i,j} v_{i,j} =\pi r_1 ^2
\left(i^{1/3} + j^{1/3} \right)^2 \left(\frac{i+j}{ij} 
\frac{8T}{\pi m_1} \right)^{1/2}.
\ee
{\color{red}Boltzmann's constant has been set to unity
(if needed, make the change $T \rightarrow k_{\rm B} T$).}
The geometric capture cross section assumed is calculated 
with the sum of the radii of the two colliding particles.
The radii are proportional to the cube root of the particle 
number, reflecting a constant density for all sizes.
{\color{red} This is obviously also a physical assumption, 
but it agrees with the experience from measurements of 
cluster ion mobilities.}
The relative speed in the square root is the average relative 
thermal speed of the particles in thermal equilibrium at 
temperature $T$, calculated as the value for a single particle 
with the reduced mass \cite{L&Lstatphys}.

The temperature in the kernels {\color{blue} thus refers} 
to the translational effective temperature.
Internal temperatures, such as those associated with 
vibrational motion, do not need to conform to this 
requirement as long as the aggregation is irreversible.
The temperature depends on time, as analyzed in detail in 
refs. \cite{BrilliantovPRE2020,OsinskiPRE2022} which 
provide equations for the time development derived from 
the classical (non-quantal) Boltzmann equation describing 
the development of energy distributions. 
Provided the translational temperature is size-independent, 
which will be assumed here, the time dependence can be 
incorporated into the kernels.
This assumption corresponds to the situations where particle 
collisions occur with {\color{blue} only a small fraction of 
fusing collisions}.
In other words, we will assume that the time development 
can be parametrized by a simple rescaling of the physical time, 
in addition to the standard scaling of the time used to cast it 
into a dimensionless form.

The parameter $q$, defined as
\be
\label{qdef}
q \equiv \pi r_1 ^2 
\left( \frac{8T}{\pi m_1} \right)^{1/2},
\ee
has dimension volume per time, and clearly depends only on 
the aggregating material and the temperature.
The latter will be time-dependent{\color{blue}, as argued,} 
and the scaled time therefore not simply proportional to 
time.
Together with the total monomer concentration it is used 
to rewrite the equations in dimensionless form.
For this purpose, it is convenient to use the quantity 
$c_0$, which is defined as the reciprocal of the volume 
that contains a single monomer, bound or not.
The scaled time is then defined as
\be
\tau \equiv t/q c_0.
\ee
This all gives the kernels the form
\be
\label{kernels}
a_{i,j} = \(i^{1/3}+j^{1/3}\)^2\(\frac{i+j}{ij}\)^{1/2}.
\ee
The scaling of the concentration means that the total 
particle number is normalized to unity:
\be
\sum_{i=1}^{\infty} Nc_N = 1.
\ee

\section{Time development of mean size}

A number of results on both mean sizes and 
scaling of distributions have been derived previously,
see e.g. refs. \cite{DongenPRL1985,LeyvrazPR2003}.
The present section serves to establish the numerical 
values of the coefficient multiplying the dependence
of {\color{blue} the mean size on} the scaled time and 
to illustrate the quality of the approximations that 
provide the coefficient by comparison with numerically 
calculated values.

The average particle size is
\be
\label{average-size2}
\olsi{N} = \frac{\sum_{N=1}^{\infty} Nc_N }
{\sum_{N=1}^{\infty} c_N} = 
\frac{1}{\sum_{N=1}^{\infty} c_N},
\ee
with the time derivative
\be
\label{smo-meansize}
\frac{\der \olsi{N}}{\der \tau} = 
-\frac{1}{\left( \sum_{N=1}^{\infty} c_N \right)^2} 
\sum_{N=1}^{\infty} \frac{\der c_N}{\der \tau} = - \olsi{N}^2 
\sum_{N=1}^{\infty} \frac{\der c_N}{\der \tau}.
\ee
For the right-hand side derivatives, the Smoluchowski 
equation is used.
Inserting it gives 
\begin{alignat}{2}
&~~~~~~~~\frac{\der \olsi{N}}{\der \tau}= \nonumber\\
-&\olsi{N}^2 \sum_{N=1}^{\infty} \left(
\sum_{i=1}^{N-1} \frac{1}{2} a_{i,N-i} c_i c_{N-i} 
- \sum_{i=1}^{\infty} a_{i,N} c_N c_i  \right).
\end{alignat}
To perform the sums, we first note that they are 
essentially identical.
This is seen from
\be
\label{doublesum}
\sum_{N=1}^{\infty} \sum_{i=1}^{N-1} a_{i,N-i} c_i c_{N-i} =
\sum_{N=1}^{\infty} \sum_{i=1}^{\infty} a_{i,N} c_i c_N.
\ee
The derivative therefore simplifies to 
\begin{alignat}{2}
\label{Nbarderiv}
\frac{\der \olsi{N}}{\der \tau} = 
\olsi{N}^2 \sum_{N=1}^{\infty}
\sum_{i=1}^{\infty} \frac{1}{2} a_{i,N} c_i c_N.
\end{alignat}
Up to this point the results hold for any set of kernels. 

The expression now needs to be approximated.
For this purpose, we note that the $a$'s are slowly 
varying functions when neither of the two indices is 
very small. 
We therefore use the approximation.
\be
\label{a-approx}
a_{i,N} \approx a_{\olsi{N},\olsi{N}} = 4\sqrt{2}\,\olsi{N}^{1/6}.
\ee
Hence 
\be
\frac{\der \olsi{N}}{\der \tau} = \frac{1}{2}4\sqrt{2}\,\olsi{N}^{1/6}
\olsi{N}^2 \sum_{N=1}^{\infty} \sum_{i=1}^{\infty} c_i c_N.
\ee
The two sums in this expression decouple and are both 
equal to $1/\olsi{N}$ by Eq. (\ref{average-size2}). 
This gives 
\be
\label{meanvalue}
\frac{\der \olsi{N}}{\der \tau} = 2\sqrt{2}\,\olsi{N}^{1/6}.
\ee
It is quite remarkable that this result is obtained 
without any knowledge about the distribution.
Only the relation Eq. (\ref{a-approx}) is required. 
Clearly, other kernels with similar properties can be 
analyzed similarly.

The time development is then simply
\be
\label{smo-mean-approx}
\olsi{N} = \left(\frac{5\sqrt{2}}{3} \tau +\tau_0 \right)^{6/5}.
\ee
{\color{red} In order to convert this into a dependence on the 
physical time, the dependence of $\tau$ on temperature is used.
The result from ref. \cite{BrilliantovPRE2020} is, with the 
notation used here and considering Eq.(\ref{average-size2}),
\be
\label{BrillT}
\frac{T}{T(0)} = \olsi{N}^{-1/3}.
\ee
This gives the linear time dependence
\be
\olsi{N} = \frac{5\sqrt{2}}{3c_0 r_1^2 
\sqrt{\frac{8\pi T(0)}{m_1}}} t + \olsi{N}(0).
\ee 
As this is close to the scaled time dependence, that 
will be used in the following.}

{\color{red} It is instructive to compare the calculated 
value in Eq.(\ref{BrillT}) with a simple estimate based on 
mean values.
Approximating the derivatives in $\d T / \d \olsi{N}$ with 
the changes in one collision of two particles of mean sizes
$\olsi{N}$, one has
\be
\frac{\d T}{\d \sum_i c_i} \approx 
\frac{-\frac{1}{4}T}{-\frac{1}{2}\sum_i c_i}
=\frac{1}{2}\frac{T}{\sum_i c_i}.
\ee
This is remarkably close to the result in Eq.(\ref{BrillT}) 
with the only difference being the replacement of 1/3 by 1/2.}

The approximate {\color{red} derivation of the time dependence 
of the mean size on time} suggests that a check with a numerical 
simulation is {\color{red} appropriate}.
\begin{figure}[ht]
\includegraphics[width=0.5\textwidth,angle=0]{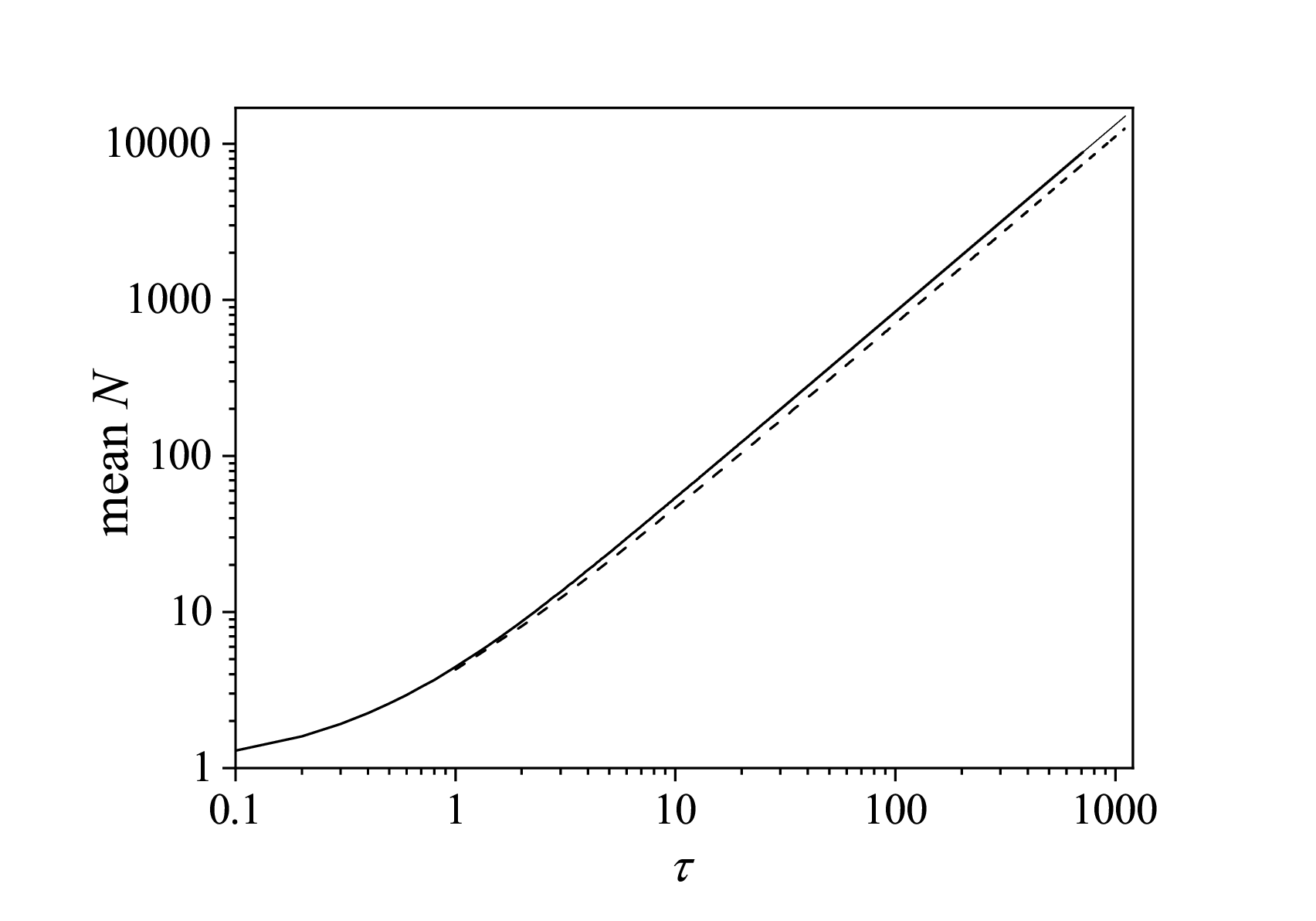}
\vspace{-0.5cm}
\begin{centering}
\caption{Numerically simulated (full line) and approximate 
(dotted line) mean particle sizes for the aggregation-only 
Smoluchowski equations.\label{simulation1}}
\end{centering}
\end{figure}
The numerical integration of the coupled differential 
equations shown in Fig. \ref{simulation1} started with 
monomers at $\tau=0$, which fixes $\tau_0$ to be 1.
The expected growth of the mean size with a power slightly 
above unity is confirmed by the simulations, and the power 
of 6/5 on the scaled time is reproduced fairly well.
The difference from the predicted value in 
Eq.(\ref{smo-mean-approx}) 
is a deviation from the predicted multiplicative factor 
of 16 \%.

Similar relations for the asymptotic forms have been 
established previously by other means (see e.g. 
\cite{DongenPRL1985}).
The agreement of the result here with those and the 
numerical calculations lends confidence in the applicability 
of Eq.(\ref{a-approx}).

\section{Scaling properties}

The scaling properties of the solutions for different 
kernels and initial conditions have been discussed extensively 
in refs. \cite{LeyvrazPR2003,FournierCMP2005}.
This section shows that the scaling is indeed 
consistent with the kernels used and that the mean sizes 
calculated previously are likewise consistent with scaling.
It also demonstrates a separation of time and $N/\olsi{N}$
variables which suggest that a partial differential equation
can be established. 
This is accomplished in the {\color{blue} section that follows}.

The kernels used in the equations here accommodate scaled 
solutions on the form
\be
\label{n-scaled}
c_N = \frac{1}{\olsi{N}^2}\tilde{c}\left(\frac{N}{\olsi{N}} \right),
\ee
where the reciprocal square of the mean size accounts for 
mass conservation and the scaling size can be taken as the 
mean size, $\olsi{N}$, without loss of generality. 

The scaling does not determine the scaled abundances 
$\tilde{c}$ \textit{per se} but provides a useful tool for 
their determination.
The rate of change of each concentration is 
\begin{alignat}{2}
\label{c-deriv0}
\color{red}{\frac{\partial c_N}{\partial \tau}}& = 
-2\frac{\dot{\olsi{N}}}{\olsi{N}\,^3}\tilde{c}\(\frac{N}{\olsi{N}} \)
- \frac{N \dot{\olsi{N}}}{\olsi{N}\,^4} \tilde{c}'\(\frac{N}{\olsi{N}} \)\nonumber \\
&= \frac{\dot{\olsi{N}}}{\olsi{N}\,^3}\left(-2 \tilde{c}\(\frac{N}{\olsi{N}} \)
- \frac{N}{\olsi{N}} \tilde{c}'\(\frac{N}{\olsi{N}} \) \right)\nonumber \\
&\equiv \frac{\dot{\olsi{N}}}{\olsi{N}\,^3}f\(\frac{N}{\olsi{N}} \),
\end{alignat}
where $\tilde{c}'$ indicates the derivative with respect 
to the argument. 
As indicated, $f$ is a function of $N/\olsi{N}$ only,
and its prefactor only of time (and the initial conditions).

The time derivative is also equal to 
\begin{eqnarray}
\label{c-deriv}
\color{red}{\frac{\partial c_N}{\partial \tau}} &=& 
\sum_{i=1}^{N-1} \frac{1}{2} a_{i,N-i} c_i c_{N-i} - 
\sum_{i=1}^{\infty} a_{i,N} c_N c_i
\\
&\approx& 
\frac{1}{2} \int_0^N a_{i,N-i} c_i c_{N-i} \der i 
-c_N \int_0^{\infty} a_{i,N} c_i \der i.
\end{eqnarray}
The kernels are homogeneous functions with exponent 1/6:
\begin{equation}
\label{a-coeff}
a_{\alpha i, \alpha j} = \alpha^{1/6} a_{i,j}.
\end{equation}
Use of this together with the scaling in Eq. (\ref{n-scaled}) 
for the concentrations allows Eq. (\ref{c-deriv}) to be 
written, with $y \equiv N/\olsi{N}$, as:
\begin{alignat}{2}
\label{c-deriv2}
&~~~~~\frac{\der c_N}{\der \tau} \approx \olsi{N}\,^{-17/6}\times\\
&\nonumber
\left[ \frac{1}{2} 
\int_0^{N/\olsi{N}} a_{\rm x,y-x} \tilde{c}_x \tilde{c}_{\rm y-x} \der x 
-\tilde{c}_y \int_0^{\infty} a_{\rm x,y} \tilde{c}_x \der x\right].
\end{alignat}
The right-hand side is a product of $\olsi{N}\,^{-17/6}$ 
and a function of $N/\olsi{N}$.
For notational simplicity, it will be denoted by $g$:
\begin{alignat}{2}
\label{c-deriv3}
\frac{\der c_N}{\der \tau} \approx \olsi{N}\,^{-17/6} 
g\(\frac{N}{\olsi{N}} \).
\end{alignat}
Equating Eqs. (\ref{c-deriv0},\,\ref{c-deriv3}) gives
\begin{equation}
\frac{\dot{\olsi{N}}}{\olsi{N}^{1/6}}
= \frac{g\(\frac{N}{\olsi{N}} \)}
{f\(\frac{N}{\olsi{N}} \)},
\end{equation}
As the left-hand side of this equation does not depend 
on $N$ and the right-hand side not on time, a separation 
of variables has been achieved.
The separation constant has already been calculated in 
Eq. (\ref{meanvalue}) to be $2\sqrt{2}$. 
Notably, the specific kernel is manifested only in the 
power 1/6.
Other kernels will give analogous results with different 
powers, provided {\color{blue} that} they are 1) slowly 
varying with size, and 2) homogeneous functions with the 
power less than unity to avoid gelation \cite{HendriksJSP1983}.

\section{Solution with the scaled abundances}

The scaling properties of the solutions will first be 
used to establish a partial differential equation for 
$\tilde{c}$. 
The partial derivative of the function with respect to 
time is, with Eq.(\ref{n-scaled}), equal to
\begin{eqnarray}
\frac{\partial c_N}{\partial \tau} = 
-2\frac{\dot{\olsi{N}}}{\olsi{N}^3}
\tilde{c}\left(\frac{N}{\olsi{N}}\right)
-\frac{N \dot{\olsi{N}}}{\olsi{N}^4}
\tilde{c}'\left(\frac{N}{\olsi{N}}\right)
\end{eqnarray}
The derivative with respect to size is
\be
\frac{\partial c_N}{\partial N} = 
\frac{1}{\olsi{N}^3} \tilde{c}'\left(\frac{N}{\olsi{N}}\right).
\ee
Substituting this equation into the previous and using 
\be
\frac{\der \olsi{N}}{\der \tau} = \frac{6}{5} \frac{\olsi{N}}{\tau} 
\ee
gives us
\begin{alignat}{2}
\label{partdiffsmolu}
&\frac{\partial c_N}{\partial \tau} = 
-\frac{6}{5\tau} \left( 2c_N + N\frac{\partial c_N}
{\partial N}\right)\nonumber\\
&~~~~~~~~~~~~~\Downarrow \nonumber\\
&\frac{\partial \ln(c_N)}{\partial \ln \tau} = 
-\frac{6}{5}\left(2 + \frac{\partial \ln(c_N)}{\partial \ln N}\right).
\end{alignat}
The coefficient 6/5 arises as $1/(1-p)$, where $p$ is the 
order of the homogeneous kernels.

Inspection shows that two types of functions solve 
the equation.
One is the pure exponential,
\be
\label{singleexp}
c_N = \olsi{N}^{-2}\e^{-N/\olsi{N}}.
\ee
The other is a {\color{red}(slightly modified)} log-normal 
function: 
\be
\label{scaled-conc}
c_N = a \olsi{N}^{\,-2} \exp\left(-\frac{1}{2s^2}\left(\ln(N) 
-\ln(N_0) \right)^2 \right).
\ee
We note that Eq. (\ref{partdiffsmolu}) also holds for 
constant kernels, provided {\color{blue} that} the factor 
6/5 is replaced by unity, consistent with it being a 
coefficient derived from the time development.

The analysis so far does not provide the criterion for 
choosing either of these two forms of solutions.
The choice is made by {\color{blue} a heuristic 
consideration of the time development of the monomer
concentration.  
From Eq. (\ref{c-deriv})} we have 
\be
\label{monomerderiv}
\frac{\d c_1}{\d \tau} = - c_1\sum_{i=1}^{\infty} a_{i,1}c_i.
\ee
{\color{blue} For size-independent kernels the time 
dependence of the monomer at long times is
\be
c_1 \propto \tau^{-2}.
\ee
A comparison with Eq. (\ref{singleexp}) identifies the 
solution for these kernels with the exponential form,
consistent with Smoluchowski's solution.
Alternatively, application of Eq. (\ref{monomerderiv}) with 
the kernels of} Eq. (\ref{kernels}) gives
\be
\label{monomerderiv2}
\frac{\d c_1}{\d \tau} &=
 - c_1\sum_{i=1}^{\infty} \(i^{1/3}+1 \)^2\(\frac{i+1}{i}\)^{1/2} c_i\\\nonumber
&\approx - c_1\sum_{i=1}^{\infty} i^{2/3} c_i
\ee
where the approximation {\color{blue} refers to} long times where 
$\langle i \rangle \gg 1$.
We approximate the sum as
\be
\sum_{i=1}^{\infty} i^{-1/3} i c_i \approx  \olsi{N}^{-1/3} 
\sum_{i=1}^{\infty} i c_i=\olsi{N}^{-1/3}.
\ee
With the known time dependence of the mean size, the 
monomer concentration {\color{blue} becomes, at long times 
where we can ignore $\tau_0$, equal to} 
\be
c_1 \propto \exp\(-\alpha \tau^{3/5}\),
\ee
with 
\be
\alpha \equiv \(\frac{5}{3}\)^{3/5}2^{-1/5} = 1.18... 
\ee
Clearly, this is not consistent with the exponential 
solution and we can therefore assign the log-normal 
distributions to the physical kernels of interest here.

The constants of integration $a, s$ and $N_0$ in 
Eq.(\ref{scaled-conc}) 
are related due to mass conservation and the known 
time dependence of the mean size.
Replacing summation with integration, mass conservation 
yields
\be
a = \frac{\olsi{N}^2}{N_0^2}
\frac{1}{s\sqrt{2\pi}}{\rm e}^{-2s^2}.
\ee
The reciprocal of the mean size is calculated similarly.
With the value of $a$ known it is calculated to
\begin{alignat}{2}
\olsi{N}^{-1} = \sum_{N=1}^{\infty} c_N = N_0^{-1} {\rm e}^{-3s^2/2}.
\end{alignat}
Hence the average size is larger than the peak value of the size distribution 
by the factor $\exp(3s^2/2)$:
\be
\frac{\olsi{N}}{N_0} =  {\rm e}^{3s^2/2}.
\ee
Inserting this into the result for $a$ gives
\be
a = \frac{1}{s\sqrt{2\pi}}{\rm e}^{s^2}.
\ee
The peak value size, $N_0$, varies with time as $\olsi{N}$, 
i.e. as $\tau^{6/5}$, {\color{blue} consistent with} the 
scaling properties of the solution.
In particular, we have that
\be
\label{log-deriv}
\frac{\dot{\olsi{N}}}{\olsi{N}} = \frac{\dot{N}_0}{N_0}.
\ee

To find the width of the distribution, represented by 
$s$, we calculate the time derivative of the peak size, 
$N_0$, with both the scaled solution containing the 
unknown $s$ and with the Smoluchowski equation.
From the scaled expression in Eq. (\ref{scaled-conc}) we 
have
\begin{eqnarray}
\label{scaledsolution}
\dot{c}_{N_0} = -2 a \frac{\dot{\olsi{N}}}{\olsi{N}} \olsi{N}^{\,-2}
= -4 \sqrt{2} a \olsi{N}^{\,-17/6},
\end{eqnarray}
where use was made of Eq. (\ref{log-deriv}) and the known 
time dependence of $\olsi{N}$.
The derivative calculated with the Smoluchowski equation is:
\be
\label{N0}
\dot{c}_{N_0} = \frac{1}{2}\, \sum_{i=1}^{N_0-1} a_{N_0-i,i} c_{N_0-i}c_i 
-c_{N_0} \sum_{i=1}^{\infty} a_{N_0,i} c_i.
\ee
In the second term, we approximate the size dependence 
of the kernels with the replacement $i \rightarrow \olsi{N}$.
The remaining sum is then known and given by $1/\olsi{N}$, 
making this term approximately equal to 
$-c_{N_0} a_{N_0,\olsi{N}}/\olsi{N}$. 
After inserting the expression for $c_{N_0}$ from the 
scaled solution, Eq. (\ref{scaled-conc}), the term becomes:
\be
-c_{N_0} \sum_{i=1}^{\infty} a_{N_0,i} c_i \approx
-a\olsi{N}^{\,-3} a_{N_0,\olsi{N}}.
\ee
The known relation between mean and peak values allows us 
to express the loss term as 
\be
\label{lossterm}
&&-c_{N_0} \sum_{i=1}^{\infty} a_{N_0,i} c_i \approx\nonumber\\
&&-a\, \olsi{N}^{-17/6} \(1 + {\rm e}^{-s^2/2}\)^2 
\(1+{\rm e}^{3s^2/2}\)^{\frac{1}{2}}.
\ee

The gain term in Eq. (\ref{N0}), which is a self-convolution 
of the abundances, is calculated with a saddle point expansion 
of the scaled solutions:
\begin{eqnarray}
\label{gainterm}
&&\frac{1}{2} \sum_{i=1}^{N_0-1} a_{N_0-i,i} c_{N_0-i}c_i 
\nonumber\\
&\approx& a\, \olsi{N}^{-17/6} 2^{-\frac{1}{6}} 
{\rm e}^{-(\ln2)^2/2s^2-3s^2/4},
\end{eqnarray}
where the value of $a$ was also used. 
Equating the two calculations of the time derivative, 
Eq. (\ref{scaledsolution}) and the sum of Eqs. (\ref{lossterm},\ref{gainterm}), 
gives two solutions for $s$.
One is $s=0$, which is physically uninteresting. 
A numerical solution for the relevant value gives $s= 0.98$.

The number gives the Full-Width-Half-Maximum of the 
distributions of 
\be
N_{\frac{1}{2}^+} - N_{\frac{1}{2}^-}
= 4.24 N_0 = 0.34\olsi{N}.
\ee
The standard deviation of the size distribution is
\be
\sigma = 1.5 \olsi{N}.
\ee
The difference between these two values reflects the 
large difference between mean and peak values:
\be
\olsi{N}\approx 4.2N_0.
\ee

\section{Comparison with numerical solutions}

A test of the scaling of the solutions of the equations 
can be made by comparing numerically calculated particle 
size distributions for different times. 
For spectra sampled at $\tau_1$ and $\tau_2$ we have that
\be
\label{spectrumscaling}
c_{N'}(\tau_2) \frac{\olsi{N}(\tau_2)^2}{\olsi{N}(\tau_1)^2}
= c_N(\tau_1),
\ee
where 
\be
\label{Nscaling}
N' \equiv N \frac{\olsi{N}(\tau_2)}{\olsi{N}(\tau_1)}
\ee

The spectra {\color{blue} used for the comparison} 
were calculated numerically with a brute force 
solution of Eq.(\ref{smo-gain}) after discretization of the 
time.
The scaled time steps used were decided in each iteration 
as 0.0005 divided by the sum over all sizes of the absolute 
rate of change, divided by the mean particle size.
This conservative value eliminated discretization errors.
Rounding errors in the double precision numbers 
were eliminated by normalization to a unit total intensity 
at every integer value of the time.
The upper limit for particle sizes included in the sums on 
the right-hand side of the equations was set to twice the 
size where the abundances dipped below $1 \times 10^{-5}$
times the highest abundance in the spectrum, but changes 
of all sizes up to $N=2\times 10^6$ were updated in each 
time step.
All sizes below the peak value were updated and included 
into the sums at each time step.

\begin{figure}[ht]
\includegraphics[width=0.35\textwidth,angle=90]{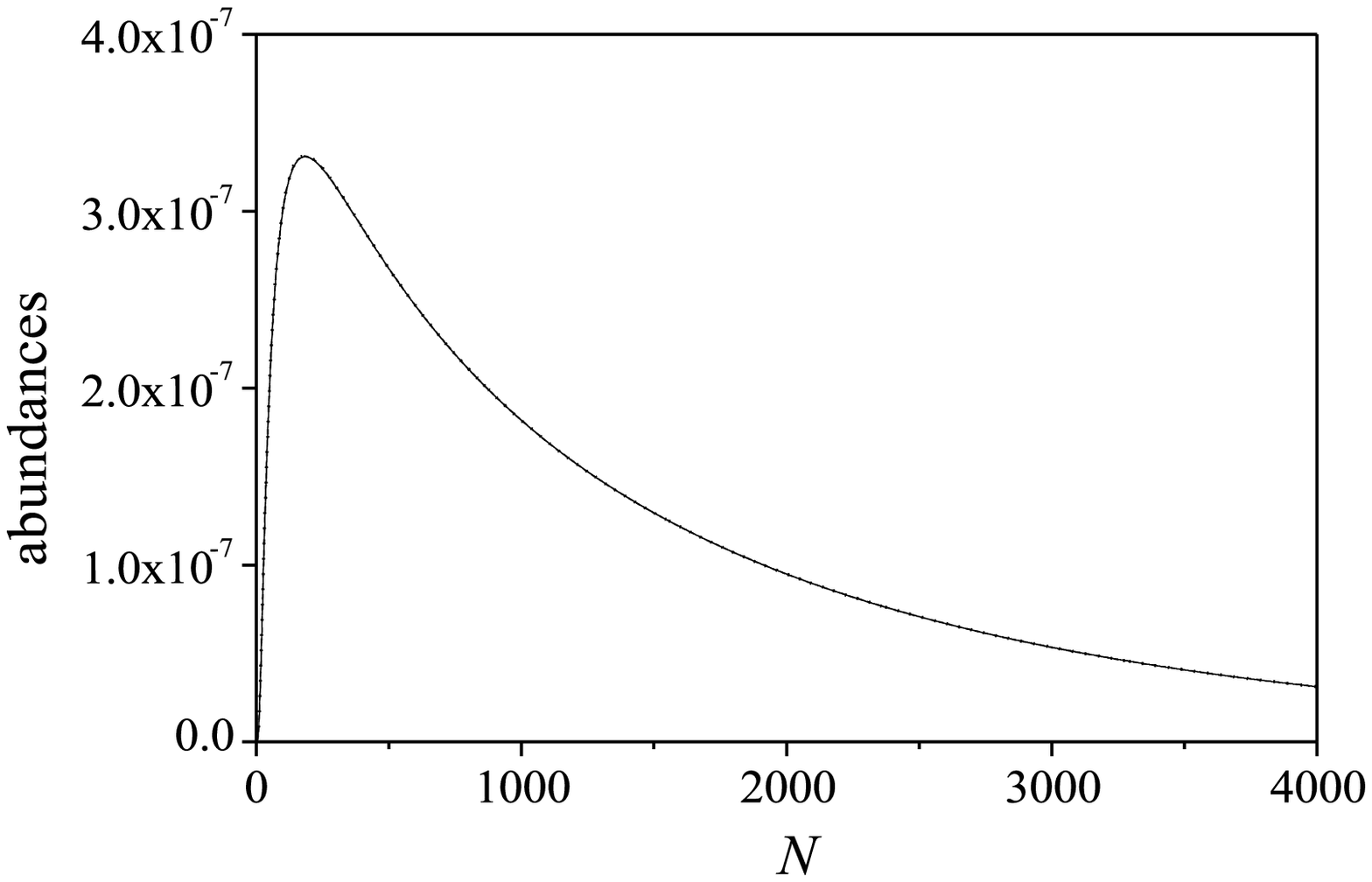}
\vspace{-0.7cm}
\begin{centering}
\caption{The particle size distribution at 
{\color{blue}$\tau=586$ and 
distributions at times 773 and 944 rescaled 
with Eqs.(\ref{spectrumscaling},\ref{Nscaling}).
The three curves overlap perfectly, indicating that the 
scaled distribution is reached below the lowest time 
plotted}.
\label{simulation2}}
\end{centering}
\end{figure}

\begin{figure}[ht]
\vspace{-0.5cm}
\includegraphics[width=0.45\textwidth,angle=0]{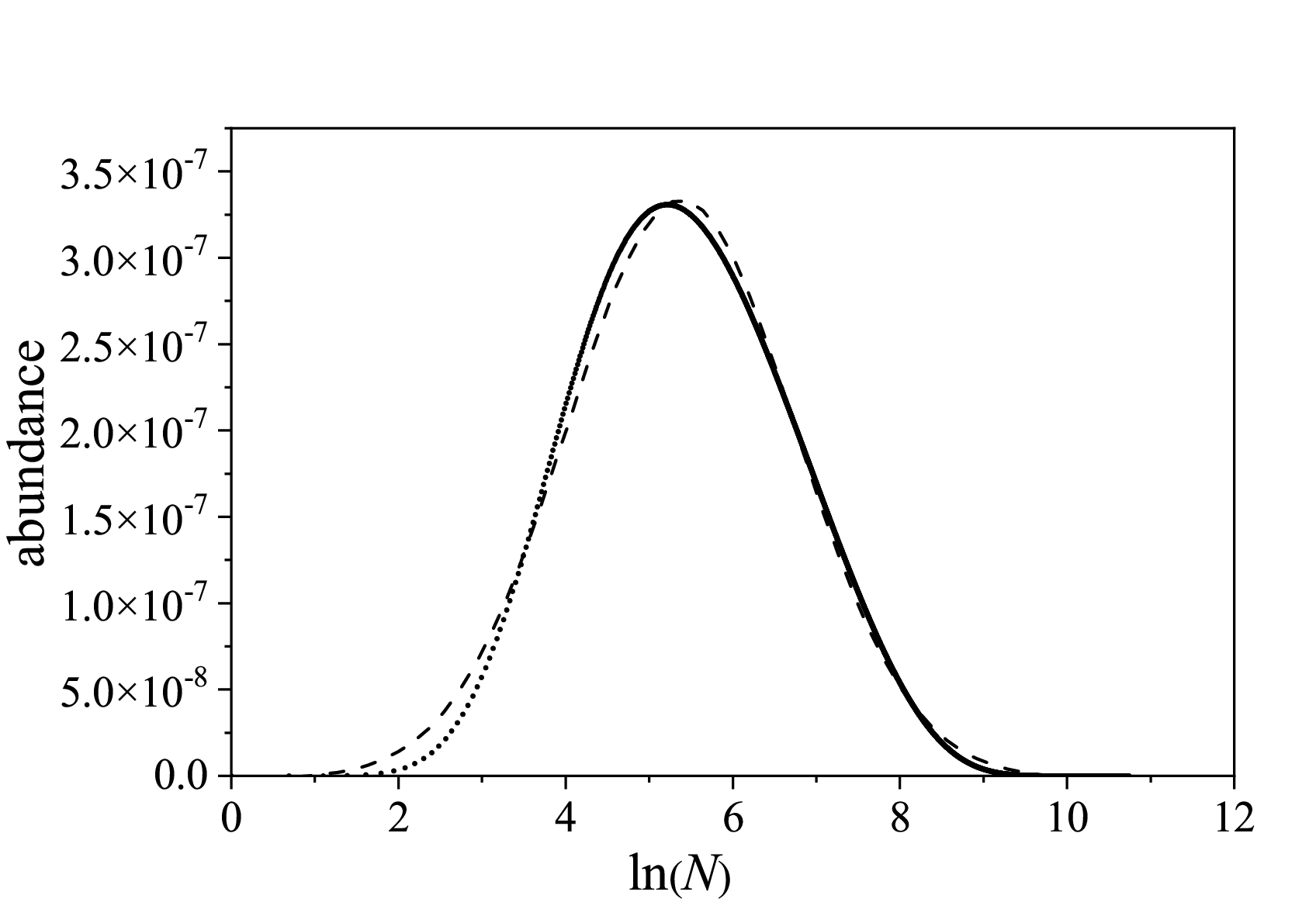}
\begin{centering}
\caption{The simulated particle size distribution at 
{\color{blue}$\tau=586$} (dots), together with {\color{blue} 
a fit of the (slightly modified) log-normal function} (dashed 
line).\label{simulation3}}
\end{centering}
\end{figure}

The distributions for {\color{blue} three} scaled times 
are shown in Fig.\ref{simulation2}.
The simulations indicate that scaling holds very 
well. 
{\color{red} Figure} \ref{simulation3} shows the comparison 
of the numerically determined distributions with a fitted 
log-normal distribution.
The fit yields the values 
$(s,\ln(N_0)) = (1.27,5.4)$, which should be compared with 
the calculated values of $(0.98,5.9)$.
As is clear from the analysis and comparison of the 
simulated data with those at longer times, the difference 
between the simulated and calculated values have reached 
their asymptotic values{\color{blue}, and} will not 
increase a longer times.

\section{Summary and discussion}

The log-normal solutions found for the kernels here agree well 
with the numerical results.
We note that the solutions in terms of log-normal functions vs. 
simple exponential size dependencies {\color{red} hinge} 
on the form of the kernels.
Furthermore, the parameters in the log-normal solutions depend 
in a tractable manner on the kernels if they are homogeneous, 
specifically on the exponent.
The approximation of the Smoluchowski equations leading to a 
partial differential equation that accommodates two fundamentally 
different solutions is of interest for the solutions of the 
special kernels found in refs. \cite{DongenJSP1988,GoodismanJCP2006} 
and for the 'free coagulation' solution in \cite{LushnikovPRE2002}.
The solutions given in the literature are, excluding a few 
of the exactly solved and gellling cases, of the form 
$N^{-\lambda}\exp\(-aN\)$ for large $N$, where 
$a$ depends on time \cite{DongenPRL1985,VillaricaJCP1993}.
In ref. \cite{VillaricaJCP1993} this is calculated by the
insertion of {\color{red} an} Ansatz into the Smoluchowski 
equations.
But as shown here, the equations in general permit two 
solutions and the single exponential is not the relevant 
one, except for the constant kernels case.
The solutions found here are approximate but  
represent fairly accurately the central part of the 
distributions where the bulk of the material is found.

The log-normal functional form is commonly applied to describe 
aggregation in cluster and nanoparticle sources.
Empirically it seems also to apply to size distributions 
generated under conditions where re-evaporation is relevant, 
beyond the condition for irreversible aggregation required 
for the present derivation.
The addition of reversibility requires the introduction of 
additional parameters and relations in the description.
With some simplifying assumptions about these, it should be 
possible to extend the approximate calculations which showed 
their usefulness in this work.
It remains to be seen how physically realistic these can be 
made.

\end{document}